\documentclass[letterpaper,twocolumn,showpacs,
prl 
,superscriptaddress
]{revtex4}

\usepackage{graphicx}
\usepackage{amssymb}
\usepackage{amsmath,amsfonts,latexsym}
\usepackage{array,tabularx,color}
\usepackage{dcolumn}               
\usepackage[normalem]{ulem}
\usepackage{comment}

\newcommand{\vect}[1]{{\mathbf #1}}
\newcommand{\Frac}[2]{\displaystyle\frac{#1}{#2}}

\begin{document}


\title{Stability and pairing in quasi-one-dimensional Bose-Fermi mixtures}

\author{Francesca M. Marchetti}
\email{francesca.marchetti@uam.es} %
\affiliation{Departamento de F\'isica Te\'orica de la Materia
Condensada, Universidad Aut\'onoma de Madrid, Madrid 28049, Spain}

\author{Th.~Jolicoeur}
\affiliation{Laboratoire de Physique Th\'eorique et Mod\`eles
  statistiques, Universit\'e Paris-Sud, 91405 Orsay, France}

\author{Meera M. Parish}
\affiliation{Princeton Center for Theoretical Science, Princeton
University, Princeton, New Jersey 08544, USA}

\date{September 5, 2009}       

\begin{abstract}
  We consider a mixture of single-component bosonic and fermionic
  atoms in an array of coupled one-dimensional ``tubes''. For an
  attractive Bose-Fermi interaction, we show that the system exhibits
  phase separation instead of the usual collapse. Moreover, above a
  critical inter-tube hopping, all first-order instabilities disappear
  in both attractive and repulsive mixtures. The possibility of
  suppressing instabilities in this system suggests a route towards
  the realization of paired phases, including a superfluid of $p$-wave
  pairs unique to the coupled-tube system, and quantum critical
  phenomena.
\end{abstract}



\maketitle

Recently, heteronuclear resonances in mixtures of bosonic and
fermionic ultracold atoms have attracted noticeable theoretical and
experimental interest, due to the possibility of generating and
exploring novel quantum phenomena in a controllable manner. For
example, by varying the interaction in a Bose-Fermi (BF) mixture,
one can, in principle, observe a quantum phase transition from a
Bose-Einstein condensate (BEC) to a normal Fermi gas phase by
binding bosons and fermions into fermionic
molecules~\cite{powell2005,marchetti2008}. Indeed, this feature has
already been 
exploited to create deeply-bound, polar
fermionic molecules~\cite{Ni2008}. However, the single biggest
impediment to realizing such novel phenomena in BF mixtures is
substantial inelastic collisions. The situation is particularly
severe on the attractive side of the heteronuclear resonance, where
a collapse of the cloud has been
observed~\cite{ospelkaus_06_3,zaccanti_06}, resulting in a sudden
loss of atoms from three-body recombination. On the repulsive side
of the resonance, an interaction-induced spatial separation of
bosons and fermions~\cite{molmer_98,viverit_00} ensures that the
atomic system is relatively stable~\cite{ospelkaus_06,modugno_07}.
However, if one sweeps through the resonance, the system once again
suffers significant inelastic losses when molecules collide with
atoms~\cite{zirbel07}.

In this Letter, we argue that many of these obstacles may be
circumvented by embedding the mixture in a two-dimensional (2D) array
of 1D tubes generated via an anisotropic optical lattice. Such a
lattice is experimentally realizable and has already been used to
explore the 1D-3D crossover in a Bose gas~\cite{stoferle2004}. While
strictly 1D BF mixtures have been investigated extensively in several
theoretical
works~\cite{lai1971,das03,cazalilla03,mathey04,akdeniz2005,imambekov2006,pollet06,mathey2007,rizzi08},
the novelty of our approach is to allow a finite hopping between
tubes, thus preserving the true long-range order of condensed phases
as found in 3D, while still maintaining the advantages of a 1D
system. In particular, 3-body recombination should be greatly reduced,
perhaps even more than in a BF mixture confined to a 3D optical
lattice (see, e.g.,~\cite{best2009}), since its rate \emph{vanishes}
for short-ranged interactions in the 1D
limit~\cite{mehta2007}. Furthermore, we demonstrate using mean-field
theory that, similarly to 1D~\cite{akdeniz2005} and contrary to
expectation~\cite{cazalilla03,rizzi08}, there is no collapse in a
quasi-1D attractive mixture. Crucially, we find that the hopping can
be used to suppress first-order instabilities in BF mixtures and, as
such, it may allow one to investigate quantum phase transitions
induced by BF pairing~\cite{powell2005,marchetti2008}, without the
intrusion of first-order transitions. In addition, we will show using
the Luttinger liquid formalism that, for a sufficiently strong BF
attraction, the coupled-tube system exhibits an exotic superfluid
phase, where $p$-wave pairing occurs between fermionic molecules
comprised of a single boson and a single fermion.

In the following, we consider a mixture of bosonic ($b$) and fermionic
($f$) atoms confined in an $N_x \times N_y$ square array of 1D tubes
of length $L_z$. We focus on the homogeneous case, but our results can
easily be mapped to the case of a harmonic trapping potential using
the local density approximation~\cite{molmer_98}.  For sufficiently
strong lattice confinement, the $xy$ motion can be approximated by a
single-band, tight-binding model (setting $\hbar=1$),
\begin{equation}
 \epsilon^{f,b}_{\vect{k}} = \Frac{k_z^2}{2m_{f,b}} + 2t
 \left[2-\cos(k_xd)-\cos(k_yd)\right] \; ,
\label{eq:dispe}
\end{equation}
where $t$ is the hopping between tubes and $d$ is the tube spacing.
Here, the transverse $xy$ momenta are restricted to the first
Brillouin zone, $|k_{x,y}| \le \pi/d$. The single-channel Hamiltonian
is thus
\begin{multline}
  \hat{H} = \sum_{\vect{k}} \left( \xi_{\vect{k}}^f
  f^{\dag}_{\vect{k}} f^{}_{\vect{k}} + \xi_{\vect{k}}^b
  b^{\dag}_{\vect{k}} b^{}_{\vect{k}} \right) + \frac{1}{L_z N_x N_y}
  \sum_{\vect{k}, \vect{k}', \vect{q}}\\
  \left[U_{BF} b^{\dag}_{\vect{k}} f^{\dag}_{\vect{k}'}
    f^{}_{\vect{k}' + \vect{q}} b^{}_{\vect{k} - \vect{q}} +
    \Frac{U_{BB}}{2} b^{\dag}_{\vect{k}} b^{\dag}_{\vect{k}'}
    b^{}_{\vect{k}' + \vect{q}} b^{}_{\vect{k} - \vect{q}}\right] \; ,
\label{eq:hamil}
\end{multline}
where $\xi^{f,b}_{\vect{k}} = \epsilon^{f,b}_{\vect{k}} - \mu_{f,b}$
and $\mu_f$ ($\mu_{b}$) is the fermionic (bosonic) chemical
potential. The contact interactions $U_{BF}$, $U_{BB}$ are effectively
1D, and we choose a repulsive boson-boson interaction $U_{BB} > 0$ to
ensure the stability of the Bose gas. Interactions between identical
fermions can be neglected due to the Pauli exclusion principle. If all
atoms experience the same transverse trapping frequency
$\omega_\perp$, then the 1D interactions, $U_{BF}$ and $U_{BB}$ can be
written simply in terms of the 3D scattering lengths $a_{BF}$ and
$a_{BB}$~\cite{olshanii1998}:
\begin{align}
 \frac{1}{U_{\alpha \beta}} = & \frac{m_{\alpha \beta}
   a_{\alpha\beta\perp}}{2}
 \left(\frac{a_{\alpha\beta\perp}}{a_{\alpha \beta}} - C \right) \; ,
\label{eq:inter}
\end{align}
where the oscillator length of the tube $a_{\alpha\beta\perp} =
\sqrt{1/m_{\alpha \beta}\omega_\perp}$ depends on the masses
$m_{BB}\equiv m_b$ and $m_{BF}=2m_fm_b/(m_f+m_b)$, while $C \simeq
1.4603/\sqrt{2}$.

The introduction of an inter-tube hopping $t$ naturally leads to a
crossover from 1D to 3D behavior. The limit $\epsilon^{f,b}_{\vect{k}}
\ll t$ recovers the isotropic 3D dispersion, while the opposite limit
$\epsilon^{f,b}_{\vect{k}} \gg 8t$ corresponds to the 1D regime. For
degenerate fermions, this implies 3D behavior when the Fermi energy
$\varepsilon_F \ll t$, i.e.\ at sufficiently small densities, and 1D
behavior when $\varepsilon_F \gg 8t$, i.e.\ at large
densities. However, for weakly-interacting degenerate bosons, the
spread of the momentum distribution is set by the temperature $T$ and
thus we require $k_B T \ll t$ and $k_B T \gg 8t$, respectively, to
access the 3D and 1D regimes. A corollary of this is that we expect
the superfluid critical temperature $T_c$ of the quasi-1D Bose gas to
be finite and scale as some positive power of $t$.  Contrast this with
the strictly 1D limit ($t=0$), where $T_c$ is strictly zero. We shall
focus on the $T=0$ limit, so the effective dimensionality will only
depend on the fermion density.

We begin by analyzing the first-order instabilities of the quasi-1D
mixture using mean-field theory. Of course, for purely 1D mixtures,
a mean-field description~\cite{das03,akdeniz2005} is unreliable
because the physics is dominated by fluctuations, and one must
instead use the Luttinger liquid formalism~\cite{cazalilla03}.
However, we expect a mean-field treatment to be reasonable for
finite inter-tube hopping, because then it works well in the
low-density 3D limit ($t/\varepsilon_F \gg 1$), as well as being
consistent with the Luttinger liquid description in the
high-density, weak-coupling, 1D regime ($8t/\varepsilon_F \ll 1$,
$|U_{BF}| \sqrt{2m_f/\varepsilon_F} \ll 1$)~\cite{cazalilla03}.
Specifically, we take $b_{\vect{k}}=\delta_{\vect{k},0} \sqrt{L_z
N_x N_y} \Phi$, so that the grand-canonical free energy density
$\Omega(\mu_f, \mu_b)=\min_{\Phi} f(\Phi, \mu_f, \mu_b)$ can be
easily evaluated by integrating out the fermionic degrees of
freedom, giving:
\begin{align}
\label{eq:freen}
  f &=-\Frac{1}{N_x N_y} \sum_{k_x, k_y}^{B.z.} \Frac{2}{3\pi}
  \frac{k_{Fz}^3}{2m_f} -\mu_b \Phi^2 + \Frac{U_{BB}}{2} \Phi^4 \; ,
  \\
 \Frac{k_{Fz}^2}{2m_f}&=\mu_f - U_{BF} \Phi^2
 -2t[2-\cos(k_xd)-\cos(k_yd)] \; .
\nonumber
\end{align}
In addition, the 1D densities of fermions and bosons in each tube are
given respectively by $n_b =\Phi^2$ and $n_f = 1/(N_x N_y) \sum_{k_x,
k_y}^{B.z.}  k_{Fz}/\pi$, so that, within mean-field, we always have a
BEC when $n_b > 0$.  Here, the system dimensionality is set by the
parameter $(\mu_f- U_{BF} \Phi^2)/t$ or, equivalently, $\pi n_f/\alpha
\sqrt{2 t m_f}$, where $\alpha =\int_0^{\pi} dk_x dk_y \sqrt{2+\cos
k_x + \cos k_y}/\pi^2 \simeq 1.35$.

\begin{figure}
\begin{center}
\includegraphics[width=1\linewidth,angle=0]{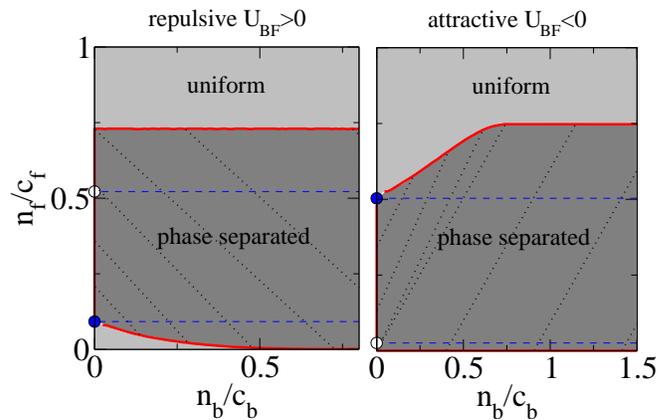}
\end{center}
\caption{(Color online) Zero temperature, mean-field phase diagrams in
  density space for a repulsive BF mixture with dimensionless hopping
  strength $t'\equiv 2m_f t/c_f^2=0.5$ (left panel), and an attractive
  mixture with $t'=0.2$ (right). Each mixture can either form a
  uniform BEC phase (light gray shaded region) or it can undergo a
  first-order transition (thick red lines) to a phase-separated state
  (dark gray). The dotted lines connect points on the first-order
  boundary with the same chemical potential.  Filled circles mark
  stable tricritical points, where $1^{\text{st}}$ and $2^{\text{nd}}$
  ($n_b=0$) order transition lines merge, while the empty circles
  correspond to unstable ones. Spinodal lines (blue dashed) divide the
  phase-separated region into the unstable domain (internal region)
  and the metastable domain (external). Note that, for the repulsive
  mixture, the phase at very low fermionic densities is always
  uniform.}
\label{fig:figu2}
\end{figure}

For a 3D, attractive ($U_{BF}<0$) mixture with no optical lattice,
it is easy to see that the free energy at large $\Phi$ is dominated
by the BF interaction term ($\propto -\Phi^5$ in 3D) and is thus not
bounded from below~\footnote{The 3D limit is recovered by expanding
Eq.~\eqref{eq:freen} in $k_{x,y}$ and setting
$2m_ftd^2=1$, so that 
$f= -k_F^5 d^2/(30
\pi^2 m_f)-\mu_b \Phi^2 + U_{BB}\Phi^4/2$, with $k_F=k_{Fz}
(t=0)$.}.  This implies that the system is unstable to collapse
at sufficiently high densities~\cite{viverit_00,chui04}. On the other
hand, in a 1D tube, the interaction term instead scales like $-\Phi^3$
at large $\Phi$ and is thus compensated by the boson-boson repulsion
($\propto \Phi^4$)~\cite{akdeniz2005}.  Therefore, contrary to what
has been previously assumed \cite{cazalilla03,rizzi08}, both 1D and
quasi-1D attractive BF mixtures will exhibit phase separation instead
of collapse.

After minimizing the free energy~\eqref{eq:freen} with respect to the
boson field $\Phi$, we can construct the phase diagram using just
three dimensionless parameters, such as the dimensionless hopping
strength $t'\equiv 2m_f t/c_f^2$, and the dimensionless densities
$n_{b,f}/c_{b,f}$, where $c_b = 2m_f|U_{BF}|^3/(\pi^2 U_{BB}^2)$, $c_f
= 2m_fU_{BF}^2/(\pi^2 U_{BB})$. The repulsive, strictly 1D ($t=0$)
case has been evaluated within mean-field in Ref.~\cite{das03}: Here,
contrary to the 3D case~\cite{viverit_00}, phase separation occurs at
\emph{low} fermionic densities, $n_f/c_f \le 3/4$, irrespective of the
boson density. Furthermore, phase separation only occurs between two
pure phases (when $n_b/c_b \le 3/4$) or between a mixed phase and a
purely bosonic phase (when $n_b/c_b > 3/4$). The topology of the
repulsive phase diagram changes substantially once $t>0$. As shown in
Fig.~\ref{fig:figu2}, a stable tricritical
point~\footnote{$\partial^3_{(\Phi^2)^3} f|_{\Phi=0} > 0$ and
$\partial_{\Phi^2} f|_{\Phi=0} = 0 = \partial^2_{(\Phi^2)^2}
f|_{\Phi=0}$.} appears at low fermionic densities and there is instead
a uniform phase for $n_f/\sqrt{2tm_f} \ll \alpha/\pi$. Here, the phase
diagram resembles the 3D phase diagram derived in
Ref.~\cite{viverit_00}, as expected.  By contrast, at higher fermionic
densities, we recover 1D behavior, such that phase separation only
exists for $n_f/c_f \lesssim \text{const}$. However, we note that
there is never phase separation between two pure phases at finite $t$,
unlike in the strictly 1D and 3D cases. Instead, phase separation
either occurs between a purely fermionic and a mixed phase (for
$n_f/c_f < 0.73$ and $n_b/c_b < 0.73$ in Fig.~\ref{fig:figu2}) or
between two mixed phases.

For the attractive case, the structure of the phase diagram does not
change when hopping is switched on. Moreover, unlike the 3D case, the
mixture displays phase separation instead of collapse, as previously
discussed. However, the phase diagram has a region where phase
separation occurs between a mixed phase and the vacuum (see
Fig.~\ref{fig:figu2}), and this may be viewed as a remnant of the
collapse in the 3D system. Note that the tricritical points have the
same values as in a repulsive mixture at the same $t'$, but their
stability is switched. Both attractive and repulsive mixtures at the
same $t'$ also feature identical spinodal lines~\footnote{At the
spinodal lines $\partial_{\Phi^2} f = 0 = \partial^2_{(\Phi^2)^2}
f=0$. For a given $t'$ the fermionic density along each spinodal line
is fixed at the density of a tricritical point, $n_f =
n_f^{\text{tcp}}$.}, which indicate when the system becomes linearly
unstable to phase separation.
\begin{figure}
\begin{center}
\includegraphics[width=0.8\linewidth,angle=0]{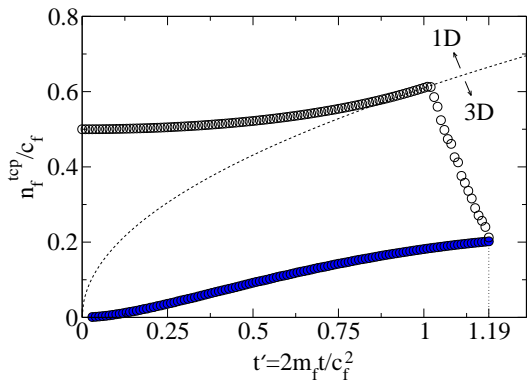}
\end{center}
\caption{(Color online) Evolution of the stable (filled blue circles)
  and unstable (empty circles) tricritical points
  $n_f^{\text{tcp}}/c_f$ ($n_b^{\text{tcp}} = 0$) as a function of the
  hopping strength $t'$ for a repulsive mixture. For attractive
  mixtures, the tricritical points have the same values, but the
  stable and unstable branches are switched. The (dashed) line
  approximately separating the 1D from the 3D regime is $n_f/c_f =
  \sqrt{2 t'} \alpha/\pi$, with $\alpha \simeq 1.35$ --- this defines
  the point, $\mu_f-U_{BF} \Phi^2=8t$, where the Fermi surface first
  touches the $xy$ band edge. Tricritical points disappear altogether
  above the critical value $t'_{\text{cr}} \simeq 1.19$.}
\label{fig:figu1}
\end{figure}

By tracking the evolution of the stable and unstable tricritical
points as a function of $t'$ (Fig.~\ref{fig:figu1}), we find that the
situation dramatically changes at larger $t'$. Notably, at the
critical value $t'_{\text{cr}} \simeq 1.19$, the stable and unstable
tricritical points merge, and the width of the phase separated region
reduces to zero. Thus, for $t'>t'_{\text{cr}}$, the system exhibits
only a uniform phase. This is a consequence of the 1D-3D crossover in
this system: eventually the instabilities of the 1D regime fall in the
low density regime where 3D behavior dominates, and vice versa.  These
results suggest that one can stabilize a BF mixture using an
appropriate 2D optical lattice. Indeed, we find that if the oscillator
length $a_{BB\perp}$ is comparable to the boson-boson scattering
$a_{BB}$, then we can have the situation where there is phase
separation or collapse in 3D and yet no instabilities in the quasi-1D
case for $t>t'_{\text{cr}}$. This is basically because the effective
1D interaction $U_{BB}$ diverges at $a_{BB\perp}= C a_{BB}$ in
Eq.~\eqref{eq:inter}.

The absence of collapse makes quasi-1D systems ideal for examining BF
pairing. In particular, the possibility of suppressing first-order
instabilities by tuning the hopping strength opens up the prospect of
investigating quantum phase transitions. For example, one could
realize a continuous quantum phase transition, where the BEC is
destroyed by the pairing. Even though our mean-field treatment does
not include the possibility of pairing, we can still get an estimate
of the instability towards BF pairing by considering the two-body
problem. In the $t = 0$ limit, the BF binding energy is approximately
$\varepsilon = m_{BF}U^2_{BF}/4$, which becomes exact in the limit
$|a_{BF}| \rightarrow 0$~\cite{bergeman2003}. However, once we switch
on the hopping, the bound state is lost when $t \simeq 0.05m_{BF}
U^2_{BF}$. Thus, if we wish to explore pairing-induced phase
transitions in the absence of first-order instabilities, we need this
``resonance'' to lie above $t'_{\text{cr}}$, i.e. we require
$U_{BB}/U_{BF} \gtrsim 0.2(1 + m_f/m_b)$.

We can determine what symmetry-broken states may exist in the quasi-1D
system by comparing the decay of different correlation functions in
the purely 1D limit. On general grounds we expect that the operator
with the slowest decay in 1D will fix the long-range ordering of the
higher-dimensional system. This is due to the fact that the exponent
$\eta$ governing the spatial decay of the operator
$\hat{\mathcal{O}}$, $\langle \hat{\mathcal{O}}_x
\hat{\mathcal{O}}_0\rangle \sim 1/x^{\eta}$, also appears in the
susceptibility as a function of temperature, $\chi(T) \sim T^{\eta
-2}$, and one can show using a mean-field approximation for the
intertube couplings that the symmetry-broken state with the most
divergent $\chi(T)$ generally has the highest $T_c$~\cite{quasi1D}.

In the strictly 1D limit, the low-energy low-wavelength effective
field theory is described by the Luttinger formalism
(bosonization)~\cite{cazalilla03}. In particular, we consider the case
where the fermionic and bosonic phase velocities are similar
$v_f=v_b=v$ and the low-energy effective Hamiltonian can be described
by introducing in- and out-of-phase phase and density fluctuations of
the mixture~\cite{cazalilla03}, $\phi_{1,2}=\frac{1}{\sqrt{2}} (\phi_b
\pm \phi_f)$ and $\theta_{1,2}=\frac{1}{\sqrt{2}} (\theta_b \pm
\theta_f)$~:
\begin{multline}
  \mathcal{H}_{\text{eff}} = \sum_{a=1,2}\Frac{v_a}{2\pi} \int dx
  \left[K_a (\partial_x \theta_a)^2 + \Frac{1}{K_a} (\partial_x
  \phi_a)^2\right] \\
  +\Frac{2U_{BF}}{(2\pi\Lambda)^2} \int dx \cos 2 \left[\sqrt{2}
  \phi_2 -\pi(n_f-n_b) x\right] \; ,
\label{eq:boson}
\end{multline}
where $K_{1,2}$ are Luttinger parameters and $\Lambda$ a cut-off.  The
in-phase mode $1$ describes a one-component (gapless) Luttinger
liquid. We emphasize that Eq.~\eqref{eq:boson} holds for any value of
the BF coupling. However, $K_{1,2}$ can only be determined
analytically in the limit of small $U_{BF}$ and one must resort to
numerics~\cite{rizzi08} when a perturbative expansion in $U_{BF}$ is
no longer accurate.

In the limit of equal filling, $n_f=n_b$, the field $\phi_2$ acquires
a gap --- the corresponding ``paired'' phase has been introduced in
Ref.~\cite{cazalilla03}. The slowest algebraic decay is then given by
the operator $\hat{\mathcal{O}} = b f$ when $K_1>2/\sqrt{3}$;
otherwise it's given by charge-density wave correlations. However
$\hat{\mathcal{O}}$ is a \emph{fermionic} operator and as such cannot
lead to condensation when we couple the tubes to access the 3D
limit. Instead we must consider the composite \emph{bosonic} operator
$\hat{\mathcal{O}}^{(n)} = f_Lf_R b^n$, whose correlations can be
evaluated from Eq.~\eqref{eq:boson}:
\begin{equation}
  \hat{\mathcal{O}}^{(n)} \propto e^{i(\sqrt{2}+
    \frac{n}{\sqrt{2}})\theta_1 +i(-\sqrt{2}+
    \frac{n}{\sqrt{2}})\theta_2}\; .
\label{eq:decay}
\end{equation}
In the paired phase the field $\phi_2$ is pinned by the relevance of
the cosine operator in Eq.~\eqref{eq:boson} and therefore the
conjugate field $\theta_2$ has exponentially decaying correlations.
As a consequence, $\hat{\mathcal{O}}^{(n)}$ also has exponentially
decaying correlations unless $n=2$. Eq.~\eqref{eq:decay} leads to a
decay law $\langle\hat{\mathcal{O}}^{(2)}_x
\hat{\mathcal{O}}^{(2)}_0\rangle\sim 1/x^{\frac{4}{K_1}}$ and
dominates over charge-density wave correlations for $K_1 >2$.
Therefore, in a system of \emph{weakly coupled tubes} we expect
condensation of the operator $\hat{\mathcal{O}}^{(2)}$, i.e.\ a
$p$-wave paired phase of fermionic molecules each comprised of a
single boson and a single fermion.
In the strictly 1D system, this operator was recognised to have
quasi-long-range order in Ref.~\cite{mathey2007}. Here, we find that,
for $K_1 >2$, the operator $\hat{\mathcal{O}}^{(2)}$ has the slowest
decaying correlations and therefore implies condensation of fermionic
molecules in the higher-dimensional system.
Moreover, the $p$-wave phase of fermionic molecules is topologically
distinct from a superfluid of $p$-wave pairs of atomic fermions
coexisting with a BEC. In the former, even though the global phase
symmetry is broken, there is a remaining subgroup $U(1)_{B-F}$ of
relative phase transformations between bosons and fermions that is
preserved, since $b^2$ and $f_L f_R$ can be rotated by opposite phase
factors without changing the superfluid order parameter. The $p$-wave
order parameter $\Delta_\vect{k}$ also breaks full spatial rotation
symmetry $SO(3)$ and, in the ground state, can either be of the form
$\Delta_{\vect{k}} \propto k_z$ (a spinless equivalent of the polar
phase of superfluid $^3$He) or $\Delta_{\vect{k}} \propto k_x+ik_y$
(the spinless variant of the $^3$He A-phase). Finally, we note that it
should be possible to access this $p$-wave phase in an attractive BF
mixture: a numerical DMRG analysis for $^{40}$K-$^{87}$Rb
mixtures~\cite{rizzi08} has shown that deep inside the paired phase
there is at least one point with $K_1\sim 2$.


\acknowledgments We are grateful to E. Burovskiy, G. Orso, and G.
Volovik for useful discussions. F.~M.~M. acknowledges financial
support from the Ram\'on y Cajal programme. Th.~J. acknowledges
support from IFRAF (Institut Francilien des Atomes Froids).

%

\end{document}